\documentclass{appolb}
\makeatletter
\@ifundefined{normalsize}{\input{size11.clo}}{}
\makeatother
\usepackage{graphicx}
\usepackage{amsmath}
\usepackage{xcolor}
\usepackage[hidelinks,hyperfootnotes=false,bookmarksnumbered=true]{hyperref}

\makeatletter
\def\Out@received{}
\makeatother
\headauthor{}
\headtitle{}

\newcommand{\muLHC}{$\mu$LHC}
\newcommand{\muTRISTAN}{$\mu$TRISTAN}
\newcommand{\LHeC}{LH$e$C}

\begin{document}

\title{LHC Based Antimuon--Nucleus Collider LH$\mu$C:\\ Nucleus Options%
\thanks{Presented at the Workshop on Ultra-cold Muons for the LHC,
Krak\'ow, Poland, 24--26 September 2026,
\texttt{https://indico.cern.ch/event/1681458/}}%
}
\author{Ali~Can~Canbay$^{1}$, Burak~Dagli$^{2}$, Arda~Goroglu$^{2}$,
Yigitali~Karaca$^{2}$, Umit~Kaya$^{1}$, Bora~Ketenoglu$^{1,\#}$,
Saleh~Sultansoy$^{2,3}$, Frank~Zimmermann$^{4}$
\address{$^1$Ankara University, Ankara, T\"urkiye\\
$^2$TOBB University of Economics and Technology, Ankara, T\"urkiye\\
$^3$Institute of Physics, Ministry of Science and Education, Baku, Azerbaijan\\
$^4$CERN, Geneva, Switzerland}
}
\maketitle
{\renewcommand{\thefootnote}{\#}%
\footnotetext{Corresponding author: \texttt{bketen@eng.ankara.edu.tr}}}

\begin{abstract}
Nucleus options of the LH$\mu$C, an antimuon--hadron collider based on the
LHC, are presented. Ultra-cold $\mu^+$ beams, based on the technology of
the J-PARC muon $g-2$/EDM experiment as proposed for \muTRISTAN, are
accelerated to 1~TeV in either a \muTRISTAN-type booster or a booster
housed in the \LHeC{} energy-recovery-linac tunnel, and collide with
LHC nucleus beams in a 3~km main ring tangential to the LHC. Collisions
with Pb, Xe, O and Ne beams provide centre-of-mass energies from 13.1 to
47.9~TeV, \emph{i.e.} about 3.3~TeV per nucleon. Luminosities and
beam--beam tune shifts are evaluated with the AloHEP software. For
$\mu$--Pb collisions, the luminosity reaches $2.7\times10^{30}$
($2.2\times10^{30}$)~cm$^{-2}$s$^{-1}$ with the \muTRISTAN-type (ERL-based)
booster, \emph{i.e.} $5.6\times10^{32}$ ($4.5\times10^{32}$)~cm$^{-2}$s$^{-1}$
per nucleon, which a Brinkmann--Dohlus dynamic focusing scheme could
increase by a factor of 2--3. The physics
potential is discussed, including an extension of the kinematic reach by
a factor of about 20 with respect to the \LHeC, access to parton
saturation at perturbative scales, nuclear parton distributions,
photon-induced processes and charged-current flavour separation.
\end{abstract}

\section{Introduction}
\label{sec:intro}

Lepton--hadron collisions have been instrumental in revealing the
structure of matter, and lepton--nucleus scattering in particular has
shown that the parton structure of nucleons is modified inside nuclei:
the EMC effect was discovered at CERN in deep inelastic muon--nucleus
scattering~\cite{EMC}, and its origin is still debated more than four
decades later. More generally, the strong interaction is responsible for
about 99\% of the mass of the visible Universe, while confinement,
hadronisation and the transitions from quarks to hadrons and to nuclei are
still far from being understood~\cite{muLHC,DagliMuA}. In nuclei, the
gluon density per unit transverse area probed at small Bjorken $x$ is
enhanced by the nuclear thickness, $\propto A^{1/3}$, so
that non-linear effects such as parton saturation are expected to set in
at harder scales than in the proton~\cite{GLR,MV,CGC}. At the same time,
the nuclear parton distributions at small $x$ remain poorly
constrained~\cite{EPPS21,nCTEQ15HQ,nNNPDF30}, which also limits the
interpretation of heavy-ion collisions at the LHC.

Exploring these phenomena requires lepton--nucleus collisions at the
highest possible energies. With nuclear beams, the Electron--Ion Collider
(EIC)~\cite{EICYR,EICwhite} will reach lepton--nucleon centre-of-mass
energies of up to about 90~GeV, the \LHeC~\cite{LHeC2021}, colliding
50~GeV electrons with the LHC ion beams, about 0.74~TeV, and a
muon--ion collider at BNL (MuIC)~\cite{MuIC,MuIC2} about 0.65~TeV. Since
the synchrotron radiation of muons is suppressed by a factor of
$(m_\mu/m_e)^4\approx2\times10^{9}$ with respect to electrons, muons can
be accelerated to the TeV scale in rings, and ring--ring muon--nucleus
colliders can reach considerably higher energies. Such colliders were
proposed by placing a muon collider ring tangential to the FCC, SppC,
LHC, HE-LHC and RHIC~\cite{DagliMuA,Acar2017,KetenogluCJP}; see
Ref.~\cite{Ketenoglu2022} for a review. However, their realisation
depends on the development of muon ionisation cooling to the level
required for a muon collider.

A different route has been opened by the ultra-cold muon technology
developed for the muon $g-2$/EDM experiment at J-PARC~\cite{gm2edm},
which provides low-emittance $\mu^+$ beams without ionisation cooling and
is the basis of the \muTRISTAN{} proposal of $\mu^+e^-$ and $\mu^+\mu^+$
colliders~\cite{muTRISTAN}. Antimuon rings of this type tangential to the
LHC, Tevatron, FCC and SppC would provide $\mu^+p$ and $\mu^+A$ collisions
at multi-TeV centre-of-mass energies~\cite{AkturkEPJP}. In particular,
the LHC-based antimuon--proton collider \muLHC~\cite{muLHC} reaches
$\sqrt{s}=5.3$~TeV with peak luminosities of $8.8\times10^{33}$ and
$7.2\times10^{33}$~cm$^{-2}$s$^{-1}$ for a \muTRISTAN-type and an
\LHeC-ERL-based booster, respectively. This concept, now developed further
under the name LH$\mu$C~\cite{LHmuC}, was discussed in detail at the
Workshop on Ultra-cold Muons for the LHC.

The LHC also offers a variety of nuclear beams. Besides Pb--Pb and
$p$--Pb collisions, Xe--Xe collisions were recorded in a pilot run in
2017~\cite{SchaumannXe}, O--O and Ne--Ne collisions were provided for the
first time in 2025~\cite{BruceONe}, and the luminosity reach of further
species is being studied for future runs~\cite{Waagaard}. An antimuon ring
tangential to the LHC can therefore exploit several nuclear species and
give access to the dependence of nuclear effects on the mass number $A$.
In this paper, we consider these nucleus options of the LH$\mu$C, referred
to below as the \muLHC-ion or $\mu A$ collider. With 1~TeV antimuons and
nuclear beams of about 2.7~TeV per nucleon, the antimuon--nucleon
centre-of-mass energy reaches about 3.3~TeV, which is about 4.5 and 5
times higher than at the \LHeC{} and the MuIC, respectively.
Section~\ref{sec:collider} describes the antimuon complex and the nucleus
beams and presents the resulting luminosities and beam--beam tune shifts
for Pb, Xe, O and Ne beams; Section~\ref{sec:physics} discusses the physics
potential, and Section~\ref{sec:concl} summarises our conclusions.

\section{The \muLHC-ion ($\mu A$) collider}
\label{sec:collider}

The proposed concept for a novel antimuon--nucleus collider framework,
which can be designated as the \muLHC-ion (or $\mu A$) collider, is built
upon the foundational principles of the recently proposed \muLHC{}
design~\cite{muLHC} and is inherently inspired by the \muTRISTAN{}
project~\cite{muTRISTAN}. The \muTRISTAN{} baseline explores the
feasibility of utilizing high-flux, ultra-cold antimuon ($\mu^{+}$) beams
derived from the J-PARC muon $g-2$/EDM experiment technologies~\cite{gm2edm}
to realize innovative $\mu^+e^-$ and $\mu^+\mu^+$ colliders. In this work,
we extend this advanced antimuon production and acceleration scheme to
forge a path toward multi-TeV scale antimuon--nucleus collisions. By
tangentially aligning a dedicated 3-km circumference antimuon main ring
with the Large Hadron Collider (LHC) tunnel, the
generated $\mu^{+}$ bunches can be brought into collision with high-energy
nucleus beams accelerated in the LHC ring.

\begin{figure}[htb]
\centerline{\includegraphics[width=10cm]{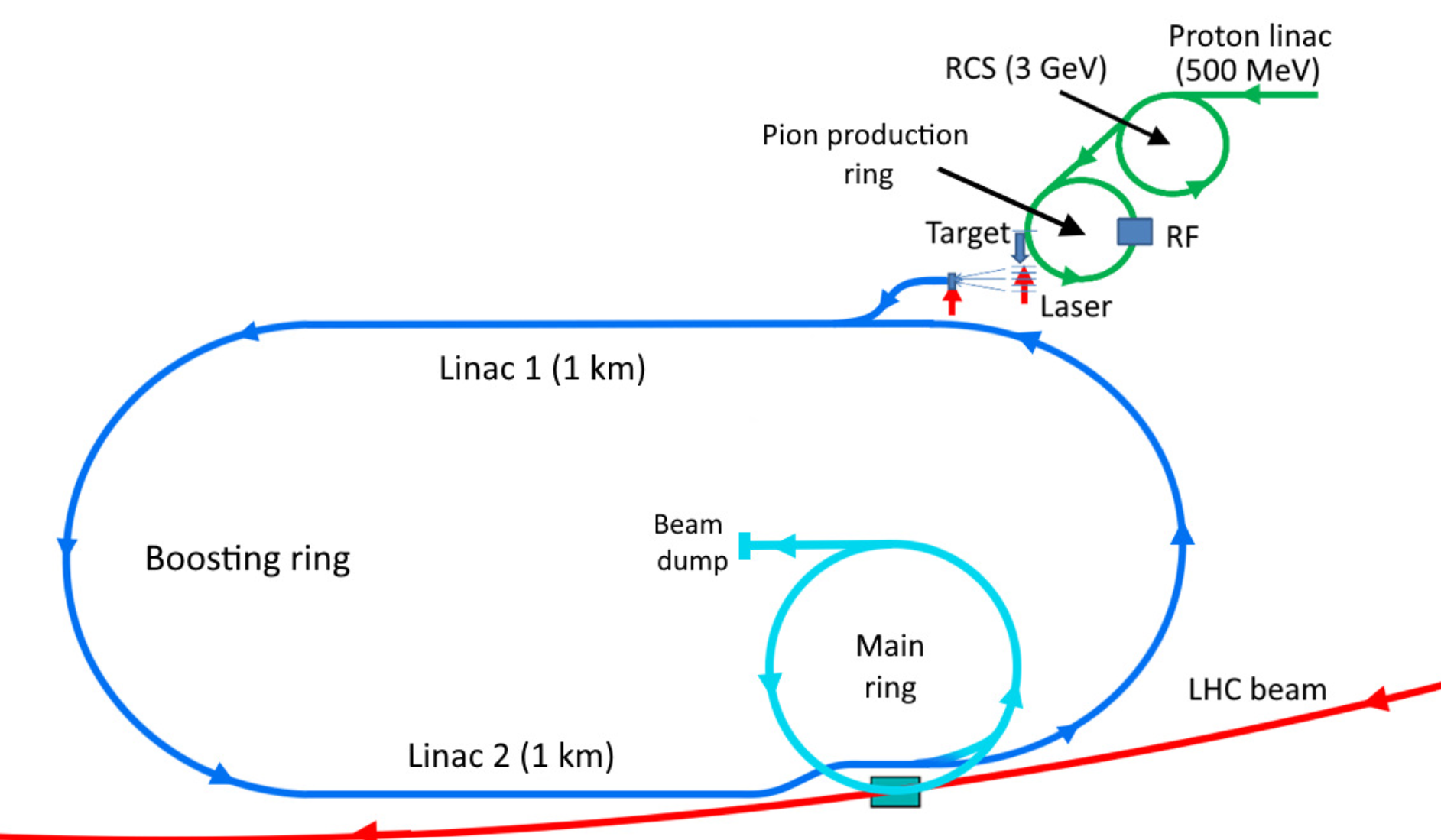}}
\caption{Schematic view of the \muLHC-ion collider with the ERL-based
booster: the booster ring and the main ring are compactly integrated
within the \LHeC{} ERL tunnel (from Ref.~\cite{muLHC}).}
\label{fig:layout}
\end{figure}

In order to accelerate the ultra-cold antimuons to the required 1~TeV
energy scale prior to injection into the main ring, we maintain the two
alternative booster ring configurations thoroughly evaluated in our
foundational \muLHC{} study~\cite{muLHC}. The first option, the
\muTRISTAN-based boosting ring, employs a large-scale racetrack geometry
consisting of two 3~km linac sections connected by arcs with a 1~km
radius. The second option offers a more compact footprint by repurposing
the tunnel infrastructure originally proposed for the Large
Hadron-electron Collider's (\LHeC) Energy Recovery Linac
(ERL)~\cite{LHeC2021}. A schematic view of this configuration, showing the
booster ring and main ring compactly integrated within the ERL tunnel in
a manner similar to that proposed for \muTRISTAN, is presented in
Fig.~\ref{fig:layout}. Regardless of the chosen boosting mechanism, the
resulting 1~TeV antimuons are injected into the tangentially aligned 3~km
main ring to collide with relativistic heavy and light nuclear species
provided by the LHC.

\begin{figure}[htb]
\centerline{%
\includegraphics[width=6cm]{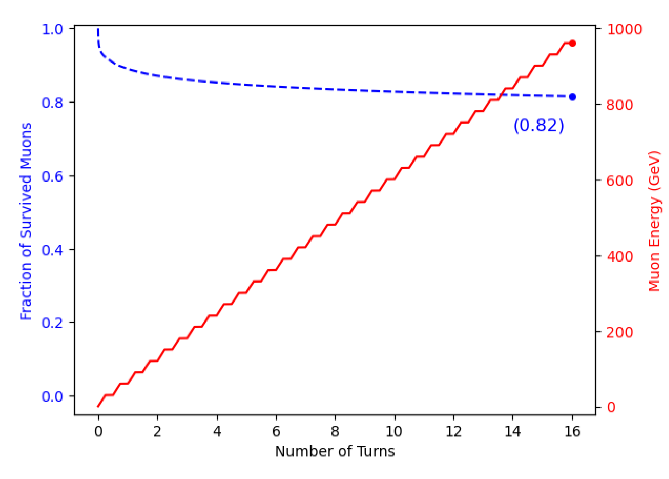}\hspace{3mm}%
\includegraphics[width=6cm]{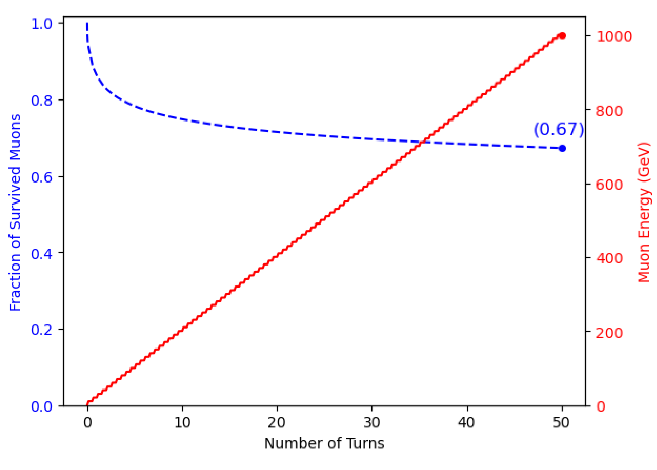}}
\caption{Fraction of surviving muons (blue, left axis) and muon energy
(red, right axis) as functions of the number of turns for the
\muTRISTAN-based (left) and ERL-based (right) booster rings.}
\label{fig:survival}
\end{figure}

Since the ultra-cold antimuon production mechanism remains identical to
the baseline \muTRISTAN{} design, the beam entering the booster ring from
the muon linac consists of 40 bunches, with each bunch containing
$4.5\times10^{10}$ muons (equivalent to 7.2~nC), at a 50~Hz repetition
rate. During the subsequent rapid acceleration phase to the 1~TeV scale,
this initial population is inevitably reduced due to natural muon decay.
Figure~\ref{fig:survival} illustrates this beam evolution, presenting the
energy gain and the fraction of surviving muons per turn side-by-side for
the \muTRISTAN-based booster (left) and the ERL-based booster (right). For
the \muTRISTAN-based configuration, accelerating the muons over 16 turns
yields a survival fraction of approximately 82\%, which means that
$3.7\times10^{10}$ muons per bunch successfully reach the main ring.
Conversely, the compact ERL-based booster requires 50 turns to achieve
the 1~TeV energy, which consequently lowers the survival fraction to
about 67\%, ultimately delivering $3.0\times10^{10}$ muons per bunch into
the main ring.

Upon injection from the booster stage, the 1~TeV ultra-cold antimuon
bunches circulate within the compact 3~km circumference main ring.
Designed to operate tangentially to the LHC infrastructure, this main
ring functions as the dedicated interaction region where the antimuons
cross paths with the LHC's relativistic nucleus beams. Achieving
high-luminosity $\mu A$ collisions in this compact setup relies heavily on
maintaining the transverse and longitudinal characteristics of the
antimuon beam. The fundamental properties and main parameters of the
$\mu^+$ beam, reflecting the precise bunch populations derived from the
survival rates of both the \muTRISTAN-based and \LHeC{} ERL-based booster
configurations, are summarized in Table~\ref{tab:muon}.

\begin{table}[htb]
\caption{Main parameters of the $\mu^+$ beam.}
\label{tab:muon}
\centerline{%
\begin{tabular}{lcc}
\hline
Parameter & \muTRISTAN-based & \LHeC-based \\
\hline
Number of muons per bunch [$10^{10}$] & 3.7 & 3.0 \\
Beam energy [GeV] & 1000 & 1000 \\
$\beta_x$ @ IP [cm] & 3.0 & 3.0 \\
$\beta_y$ @ IP [cm] & 0.7 & 0.7 \\
Bunch length [mm] & 2 & 2 \\
Norm. emittance [$\mu$m] & 4 & 4 \\
Number of bunches per ring & 40 & 40 \\
Collision frequency [MHz] & 4 & 4 \\
Circumference [km] & 3 & 3 \\
\hline
\end{tabular}}
\end{table}

The proposed $\mu A$ collider will utilize the various ion beams already
operated at the LHC, which are lead (Pb), xenon (Xe), oxygen (O) and
neon (Ne). The baseline parameters for the Pb beam are adopted directly
from Table~4 of the \LHeC{} 2021 CDR~\cite{LHeC2021}. For Xe, O and Ne
beams, the operational parameters are derived from the Particle Data
Group (PDG) records~\cite{PDG2026}. Collisions of all three species have
already been performed at the LHC, at the energies per nucleon listed in
Table~\ref{tab:ion}: Xe--Xe collisions in a pilot run in
2017~\cite{SchaumannXe}, and O--O and Ne--Ne collisions, for the first
time, in a special run in 2025~\cite{BruceONe}.
However, to maximize the luminosity, the beta function at the
interaction point for these ions is set to 10~cm, mirroring the \LHeC{}
CDR parameters for lead ions. The main parameters of these nucleus beams
are detailed in Table~\ref{tab:ion}.

\begin{table}[htb]
\caption{Main parameters of the nucleus beams.}
\label{tab:ion}
\centerline{%
\begin{tabular}{lcccc}
\hline
Parameter & Pb & Xe & O & Ne \\
\hline
Number of particles per bunch [$10^{10}$] & 0.018 & 0.027 & 0.29 & 0.20 \\
Beam energy [TeV] & 574 & 350.88 & 42.88 & 53.60 \\
Energy per nucleon [TeV] & 2.76 & 2.72 & 2.68 & 2.68 \\
$\beta$ function @ IP [m] & 0.10 & 0.10 & 0.10 & 0.10 \\
Normalized emittance [mm$\cdot$mrad] & 1.5 & 0.87 & 1.3 & 1.77 \\
Transverse emittance [nm] & 0.51 & 0.30 & 0.46 & 0.62 \\
Number of bunches per ring & 1200 & 16 & 62 & 62 \\
Rev. frequency [Hz] & 11245 & 11245 & 11245 & 11245 \\
Circumference [km] & 26.659 & 26.659 & 26.659 & 26.659 \\
\hline
\end{tabular}}
\end{table}

Based on the defined antimuon and nucleus beam parameters, the
interaction region dynamics, including the luminosity ($L$), averaged
over the muon decay between successive injections, and the beam--beam
tune shift ($\xi$), were simulated with matched beam sizes at the IP
using the AloHEP software~\cite{AloHEP}. The fundamental characteristics of the resulting
multi-TeV scale $\mu A$ collisions, considering both the
\muTRISTAN-based and ERL-based booster options, are summarized in
Table~\ref{tab:muA}.

\begin{table}[htb]
\caption{Main parameters of the $\mu A$ colliders. The centre-of-mass
energy and luminosity per nucleon are $\sqrt{s_{\mu N}}=2\sqrt{E_\mu E_N}$
and $L_{\mu N}=A\,L$, respectively.}
\label{tab:muA}
\centerline{%
\begin{tabular}{lcccc|cccc}
\hline
 & \multicolumn{4}{c|}{\muTRISTAN-based} & \multicolumn{4}{c}{\LHeC-based} \\
Parameter & $\mu$-Pb & $\mu$-Xe & $\mu$-O & $\mu$-Ne
          & $\mu$-Pb & $\mu$-Xe & $\mu$-O & $\mu$-Ne \\
\hline
$\sqrt{s}$ [TeV] & 47.9 & 37.5 & 13.1 & 14.6 & 47.9 & 37.5 & 13.1 & 14.6 \\
$\sqrt{s}$ per nucleon [TeV] & 3.32 & 3.30 & 3.27 & 3.27 & 3.32 & 3.30 & 3.27 & 3.27 \\
$L$ [$10^{30}$\,cm$^{-2}$s$^{-1}$] & 2.67 & 0.306 & 8.30 & 4.25 & 2.16 & 0.248 & 6.73 & 3.44 \\
$L$ per nucleon [$10^{32}$\,cm$^{-2}$s$^{-1}$] & 5.55 & 0.395 & 1.33 & 0.850 & 4.49 & 0.320 & 1.08 & 0.688 \\
$\sigma_{x,y}$ [$\mu$m] & 7.14 & 5.48 & 6.78 & 7.87 & 7.14 & 5.48 & 6.78 & 7.87 \\
$\xi_\mu$ [$10^{-3}$] & 4.00 & 3.95 & 6.29 & 5.42 & 4.00 & 3.95 & 6.29 & 5.42 \\
$\xi_A$ [$10^{-3}$] & 1.19 & 2.17 & 1.72 & 1.28 & 0.96 & 1.76 & 1.40 & 1.04 \\
\hline
\end{tabular}}
\end{table}

For the Xe, O and Ne options, the luminosity is limited mainly by the
small numbers of bunches in the corresponding LHC runs, 16 and 62
(Table~\ref{tab:ion}), which reduce the collision frequency to 0.18 and
0.70~MHz, respectively, instead of 4~MHz. Filling schemes similar to that
of lead, as considered for future LHC runs with various ion
species~\cite{Waagaard}, would increase these luminosities by factors of
about 22 for Xe and 6 for O and Ne. In addition, the very short
ultra-cold antimuon bunches make Brinkmann--Dohlus dynamic
focusing~\cite{BrinkmannDohlus} at the interaction region possible, which
could increase the luminosity by a further factor of 2--3 and
significantly enhance the discovery potential of the $\mu A$ collider.

\section{Physics potential of $\mu^{+}A$ collisions}
\label{sec:physics}

\subsection{Kinematic reach}

In deep inelastic lepton--nucleus scattering the accessible region of the
$(x,Q^2)$ plane is fixed by the centre-of-mass energy through
\begin{equation}
Q^2 = x\,y\,s\,, \qquad s = 4E_\mu E_N\,,
\end{equation}
where $y$ denotes the inelasticity and $E_N$ the beam energy per nucleon.
Because the energy per nucleon of the LHC ion beams varies by less than
3\% across the species considered here, the centre-of-mass energy per
nucleon is common to all of them to good accuracy,
$\sqrt{s_{\mu N}} = 3.27$--$3.32$~TeV. The choice of ion therefore does not
alter the kinematic domain that is opened. What it changes is the
magnitude of the nuclear enhancement factors, to which we return in
Sect.~\ref{sec:sat}.

\begin{figure}[htb]
\centerline{\includegraphics[width=11cm]{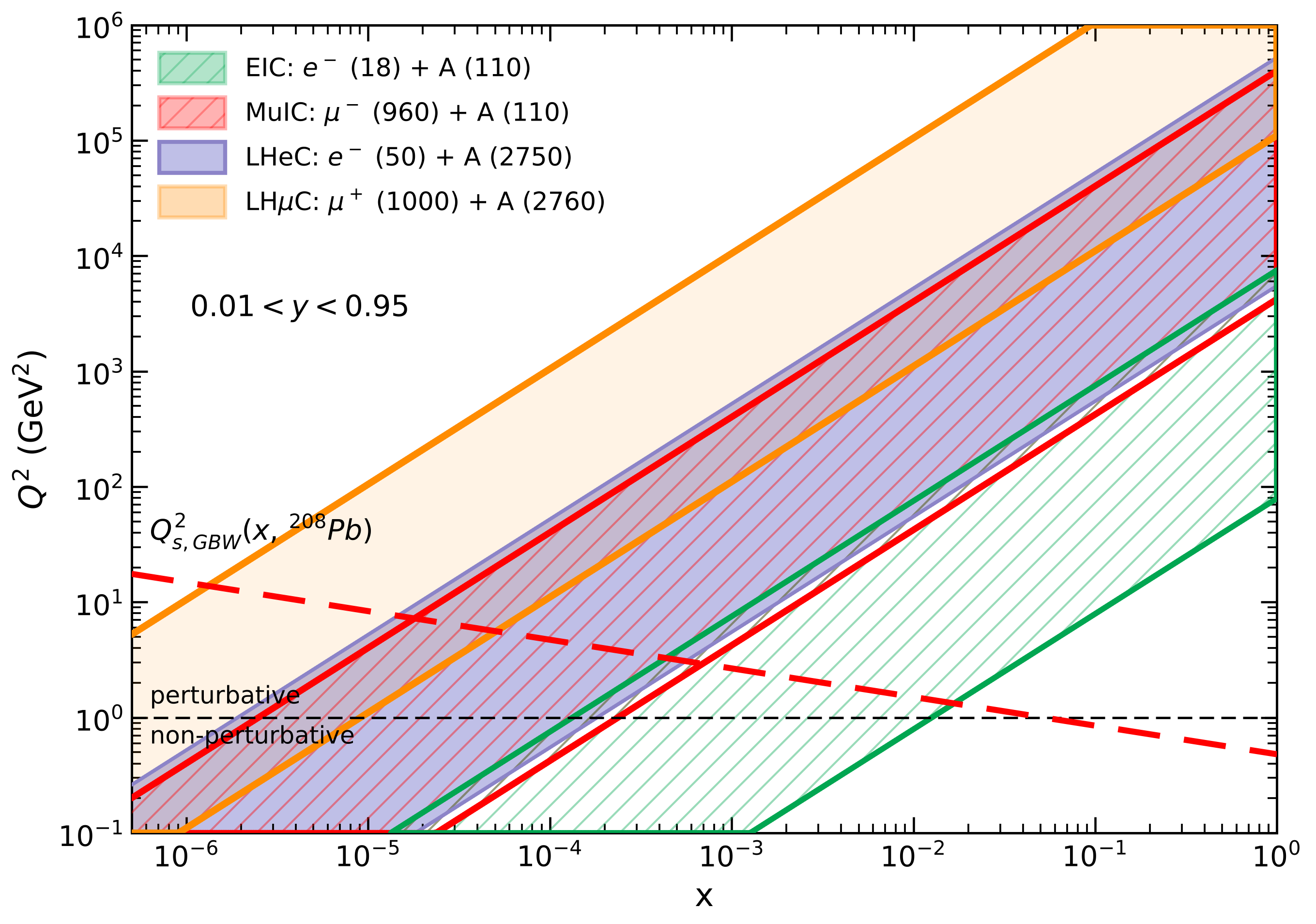}}
\caption{Kinematic coverage in the $(x,Q^2)$ plane for deep inelastic
lepton--nucleus scattering at the EIC, the MuIC, the \LHeC{} and the
proposed $\mu^{+}A$ collider, for $0.01 < y < 0.95$. Beam energies are
given in GeV and, for the nuclear beams, per nucleon. The dashed curve
indicates the saturation scale of the lead nucleus in the GBW
model~\cite{GBW}, with the parameters of Ref.~\cite{GBW2018}. A comparison of this type, for lower-energy
configurations, is presented in Ref.~\cite{MuIC}.}
\label{fig:kinA}
\end{figure}

Figure~\ref{fig:kinA} shows the region covered by the $\mu^{+}A$ option
for $0.01 < y < 0.95$, evaluated for lead, together with the corresponding
regions of the EIC~\cite{EICYR,EICwhite}, the MuIC~\cite{MuIC} and the
\LHeC~\cite{LHeC2021} in its nuclear configuration. At the kinematic limit
the smallest Bjorken $x$ accessible at $Q^2 = 10$~GeV$^2$ is
$9.1\times10^{-7}$, while at $x = 10^{-4}$ the reach in $Q^2$ extends to
$1.1\times10^{3}$~GeV$^2$. Both values exceed those of the \LHeC{} in its
nuclear configuration by about a factor of 20, and those of the MuIC by
close to a factor of thirty. Relative to $\mu^{+}p$
operation~\cite{AkturkEPJP}, for which the corresponding limit is
$3.6\times10^{-7}$, the lower energy per nucleon of the ion beams raises
the smallest accessible $x$ by a factor of 2.5. As shown below, this is
not the relevant figure of merit, since the physics of the small-$x$
region is governed by the position of $x$ relative to the nuclear
saturation scale rather than by its absolute value.

\subsection{Parton saturation}
\label{sec:sat}

At small $x$ the rapid growth of the gluon density is eventually tamed by
gluon recombination~\cite{GLR}, and the evolution becomes non-linear, as
described by the Colour Glass Condensate~\cite{MV,CGC}. The onset of this
regime is characterised by the saturation scale, which in the GBW
model~\cite{GBW} reads
\begin{equation}
Q_s^2(x,A) = Q_0^2\left(\frac{x_0}{x}\right)^{\lambda} A^{1/3}\,,
\end{equation}
with $Q_0^2 = 1$~GeV$^2$, $x_0 = 4.0\times10^{-5}$ and $\lambda = 0.248$ from
the fit to the final HERA data including heavy quarks~\cite{GBW2018}.
The nuclear enhancement factor $A^{1/3}$, which reflects the nuclear
thickness~\cite{KowalskiTeaney}, ranges from 2.52 for oxygen to 5.92 for
lead, so that a nucleus saturates at a considerably harder scale than a
proton at the same $x$.

Table~\ref{tab:sat} gives, for each species, the interval in $x$ in which
the saturation scale lies inside the acceptance, \emph{i.e.}
$Q_s^2(x,A) = y\,s\,x$ with $0.01 < y < 0.95$. For lead this interval is
$1.3\times10^{-6} < x < 5.1\times10^{-5}$, with $Q_s^2$ decreasing from
13.8 to 5.6~GeV$^2$, so that the transition to the saturated regime can be
crossed by varying $Q^2$ at fixed $x$ in the perturbative domain. With
protons, the same antimuon beam would reach the saturation scale only at
the edge of the accessible region, at $x = 1.5\times10^{-7}$ and
$Q_s^2 = 4.0$~GeV$^2$. For lead and xenon $Q_s^2$ stays at about
5~GeV$^2$ or above across the whole window, whereas for oxygen and neon
it falls to about 3~GeV$^2$ at its upper end; lead is therefore the reference species for
the saturation programme, with xenon as a close alternative.

\begin{table}[htb]
\caption{Nuclear enhancement factor, saturation scale at a representative
value of $x$, and the interval in $x$ within which the saturation scale
falls inside the acceptance $0.01 < y < 0.95$, for the four ion species
considered.}
\label{tab:sat}
\centerline{%
\begin{tabular}{lcccc}
\hline
Species & $A^{1/3}$ & $Q_s^2$ at $x=10^{-5}$ & Saturation window & $Q_s^2$ across \\
 & & [GeV$^2$] & in $x$ & window [GeV$^2$] \\
\hline
Pb & 5.92 & 8.4 & $1.3\times10^{-6}$--$5.1\times10^{-5}$ & 13.8--5.6 \\
Xe & 5.05 & 7.1 & $1.2\times10^{-6}$--$4.5\times10^{-5}$ & 12.1--4.9 \\
Ne & 2.71 & 3.8 & $7.2\times10^{-7}$--$2.8\times10^{-5}$ & 7.3--3.0 \\
O  & 2.52 & 3.6 & $6.8\times10^{-7}$--$2.6\times10^{-5}$ & 6.9--2.8 \\
\hline
\end{tabular}}
\end{table}

For lead, the luminosity of the \muTRISTAN-based option
(Sect.~\ref{sec:collider}) corresponds to about 5.5~fb$^{-1}$ per nucleon
and per year ($10^7$~s). A leading-order estimate including a nuclear
suppression factor of 0.7 then gives of order $10^{9}$ deep inelastic
events per year in the saturation window with $5 < Q^2 < 50$~GeV$^2$, so
that the measurements would be limited by systematic rather than
statistical uncertainties. Complementary probes of the saturation dynamics
are the structure functions $F_2^A$ and $F_L^A$, diffractive deep
inelastic scattering~\cite{GBW} and exclusive vector-meson production,
which also images the transverse distribution of gluons in the
nucleus~\cite{KowalskiTeaney}.

\subsection{Nuclear parton distributions}

Nuclear parton distributions are determined far less precisely than those
of the free proton. In the global analyses currently available, such as
EPPS21~\cite{EPPS21}, nCTEQ15HQ~\cite{nCTEQ15HQ} and
nNNPDF3.0~\cite{nNNPDF30}, the gluon density at small $x$ carries by a
wide margin the largest uncertainty. The origin of this situation is
partly kinematic. Existing constraints come from fixed-target experiments
confined to large $x$, where the nuclear modification of structure
functions was first established~\cite{EMC}, and from hadronic collisions
in which nuclear and partonic effects are entangled, leaving the small-$x$
region essentially unconstrained by direct measurement.

The range covered in Fig.~\ref{fig:kinA} addresses this region directly.
Measurements of $F_2^A$ would span shadowing, anti-shadowing and the EMC
effect within a single experiment, avoiding the systematic differences
that arise when datasets taken with different targets and techniques are
combined. The wide lever arm in $Q^2$ at fixed $x$ additionally allows the
nuclear gluon density to be extracted from scaling violations, providing a
determination independent of the direct measurement through $F_L^A$,
while open charm and beauty production constrain the same quantity at
scales set by the heavy quark mass. Improved nuclear parton distributions
are, in addition, a prerequisite for the interpretation of heavy-ion
collisions at the LHC, where initial-state and final-state medium effects
can only be separated if the former are independently known.

\subsection{Photon-induced processes and searches for new physics}
\label{sec:bsm}

When the momentum transfer is small enough that the nucleus recoils as a
whole, the exchanged photon couples coherently to all $Z$ protons and the
cross section acquires the enhancement
\begin{equation}
\sigma_{\rm coh} \propto Z^2\,,
\end{equation}
which for lead corresponds to a factor of 6724 with respect to a proton
target. Together with the flux of quasi-real photons radiated by the
antimuon beam, described by the equivalent photon
approximation~\cite{Budnev}, this makes the $\mu^{+}A$ collider an
effective $\gamma A$ facility operating alongside its deep inelastic
programme.

The quadratic dependence on $Z$ is what renders such a facility
competitive in searches for weakly coupled states, since the smallness of
the coupling is partly compensated by the coherent flux. A systematic
study of this potential has recently been carried out for a muon--ion
collider at lower energy~\cite{MuSIC}, covering axion-like particles
coupling to photons and produced through the Primakoff mechanism,
muonphilic vector bosons such as those associated with an
$L_\mu - L_\tau$ gauge symmetry~\cite{HeZprime}, lepton-flavour violating
leptoquarks~\cite{Dorsner} and heavy sterile neutrinos. The muonic initial
state is central to several of these channels, since it couples directly
to the currents in question, a feature that electron machines cannot
reproduce. Lepton-flavour violating transitions of the type
$\mu^{+}A \to \tau^{+}A$ benefit from the same coherent enhancement and
probe couplings that are difficult to isolate at hadron colliders, where
they are obscured by large QCD backgrounds. Given that the centre-of-mass
energy per nucleon considered here exceeds that of Ref.~\cite{MuSIC} by
roughly a factor of five, a corresponding extension of the accessible
parameter space may be anticipated.

\subsection{Charged-current flavour separation}

A further feature follows from the charge of the lepton beam, which in
the present proposal consists of antimuons produced by the ultra-cold
muon technique underlying the \muTRISTAN{} concept~\cite{muTRISTAN}. The
charged-current process
\begin{equation}
\mu^{+}A \to \bar{\nu}_\mu X
\end{equation}
proceeds through $W^{+}$ exchange and therefore selects a different
combination of quark flavours than a $\mu^{-}$ beam would, giving access
to the $d$ and $s$ content of the nuclear sea. Comparing neutral- and
charged-current cross sections at the same $x$ and $Q^2$ would thus allow
flavour-dependent nuclear modifications to be measured, in the same
kinematic region in which the saturation and nuclear parton distribution
programmes are carried out.

Quantitative projections for these channels, including the impact of
pseudodata on global nuclear parton distribution fits and the reach in
the parameter space of the new physics scenarios discussed in
Sect.~\ref{sec:bsm}, require a detector simulation and are left to
forthcoming work.

\section{Conclusions}
\label{sec:concl}

We have considered the nucleus options of the LH$\mu$C, in which
ultra-cold antimuons, accelerated to 1~TeV, collide with the LHC ion
beams in a 3~km main ring tangential to the LHC. The antimuon complex is
the same as for the $\mu^+p$ option: either a \muTRISTAN-type booster or a
booster re-using the \LHeC{} ERL tunnel delivers $3.7\times10^{10}$ or
$3.0\times10^{10}$ antimuons per bunch to the main ring. Collisions with
Pb, Xe, O and Ne beams, all of which have already been provided by the
LHC, reach centre-of-mass energies of 13.1--47.9~TeV, corresponding to
about 3.3~TeV per nucleon.

For $\mu$--Pb collisions, the luminosity reaches
$2.7\times10^{30}$ ($2.2\times10^{30}$)~cm$^{-2}$s$^{-1}$ with the
\muTRISTAN-type (ERL-based) booster, \emph{i.e.} $5.6\times10^{32}$
($4.5\times10^{32}$)~cm$^{-2}$s$^{-1}$ per nucleon.
Brinkmann--Dohlus dynamic focusing, made possible by the very
short antimuon bunches, could increase the luminosity by a factor of 2--3,
and LHC filling schemes with more ion bunches would substantially raise
the luminosities of the Xe, O and Ne options.

The physics case is compelling. The kinematic reach exceeds that of the
\LHeC{} in its nuclear configuration by about a factor of 20. For Pb and
Xe, the saturation scale crosses the acceptance at
$Q_s^2\approx5$--14~GeV$^2$, so that the onset of non-linear QCD dynamics
can be studied in the perturbative regime, while lighter nuclei probe its
dependence on $A$. The same measurements constrain the small-$x$ nuclear
gluon density through several independent observables, the coherent $Z^2$
enhancement makes the collider an effective $\gamma A$ facility for
searches for new physics, and the antimuon beam allows charged-current
measurements to separate the quark flavours of the nuclear sea.

In view of the very broad physics potential of the nucleus options of the
LH$\mu$C, we invite the nuclear physics community, and in particular the
Nuclear Physics European Collaboration Committee (NuPECC), to investigate
this potential in detail.

\section*{Acknowledgements}

We thank K.~Piotrzkowski and the Krak\'ow group for organising this workshop. This work is partially supported by T\"UB\.ITAK under grant No.~125F120.

\bibliographystyle{unsrt}
\bibliography{refs}

\end{document}